\documentclass[draft,grl]{agutex}
   \usepackage[dvips]{graphicx}
\usepackage{ulem}

\authorrunninghead{MARISALDI ET AL.}

\titlerunninghead{AGILE ENHANCED TGF DETECTION}

\authoraddr{M. Marisaldi,
INAF-IASF Bologna, Via Gobetti 101, I-40129 Bologna, Italy
(marisaldi@iasfbo.inaf.it)}

\begin{document}

%% ------------------------------------------------------------------------ %%
%
%  TITLE
%
%% ------------------------------------------------------------------------ %%

\title{Enhanced detection of Terrestrial Gamma-Ray Flashes by AGILE}

%% ------------------------------------------------------------------------ %%
%
%  AUTHORS AND AFFILIATIONS - 2 methods
%
%% ------------------------------------------------------------------------ %%

% Method 1 
% For three or fewer author/affiliation blocks, use \author{} and \affil{}

% \author{R. C. Bales}
% \affil{Department of Hydrology and Water Resources,
% University of Arizona, Tucson, Arizona, USA}

% \author{E. Mosley-Thompson}
% \affil{Department of Geography, Ohio State University,
% Columbus, Ohio, USA}

% \author{J. R. McConnell}
% \affil{Desert Research Institute, Division of Hydrologic Sciences,
% Reno, Nevada, USA}

% ---------------
% Method 2 
% For more than three author/affiliation blocks,
% use \author{\altaffilmark{}} and \altaffiltext{}
% \altaffilmark will produce footnote;
% matching altaffiltext will appear at bottom of page.
% May use \\ to start a new line.

\authors{M.~Marisaldi, \altaffilmark{1,2}
A.~Argan,  \altaffilmark{3}
A.~Ursi,  \altaffilmark{4,3}
T.~Gjesteland, \altaffilmark{5,2}
F.~Fuschino, \altaffilmark{6,1}  % <---
C.~Labanti, \altaffilmark{1}
M.~Galli, \altaffilmark{7}
M.~Tavani,  \altaffilmark{3,4}
% C.~Price, \altaffilmark{6}
C.~Pittori,  \altaffilmark{8,9}
F.~Verrecchia,  \altaffilmark{8,9}
F.~D'Amico,  \altaffilmark{10}
N.~\O stgaard,  \altaffilmark{2}
S.~Mereghetti,  \altaffilmark{11}
R.~Campana, \altaffilmark{1}
P.W.~Cattaneo,  \altaffilmark{12}
A.~Bulgarelli,  \altaffilmark{1}
S.~Colafrancesco,  \altaffilmark{9,13}
S.~Dietrich, \altaffilmark{14}
 F.~Longo, \altaffilmark{15,16}
% E.~Del~Monte,  \altaffilmark{3}
% G.~Barbiellini,  \altaffilmark{14,15}
% A.~Giuliani,  \altaffilmark{9}
% A.~Chen,  \altaffilmark{9,12}
F.~Gianotti,  \altaffilmark{1}
P.~Giommi,  \altaffilmark{8}
% F.~Lazzarotto,  \altaffilmark{3}
% A.~Morselli,  \altaffilmark{16}
% M.~Rapisarda,  \altaffilmark{17}
A.~Rappoldi,  \altaffilmark{12}
M.~Trifoglio,  \altaffilmark{1}
A.~Trois,  \altaffilmark{17}
% S.~Vercellone,  \altaffilmark{19}
}

% E.~Moretti,  \altaffilmark{3}
% E.~Costa,  \altaffilmark{4}
% S.~Cutini,  \altaffilmark{9}
% I.~Donnarumma,  \altaffilmark{4}
% Y.~Evangelista,  \altaffilmark{4}
% M.~Feroci,  \altaffilmark{4}
% I.~Lapshov,  \altaffilmark{4,21}
% P.~Lipari,  \altaffilmark{10,11}
% L.~Pacciani,  \altaffilmark{4}
% P.~Soffitta,  \altaffilmark{4}
% F.~Boffelli,  \altaffilmark{13}
% P.~Caraveo,  \altaffilmark{7}
% V.~Cocco,  \altaffilmark{4}
% F.~D'Ammando,  \altaffilmark{4,6}
% G.~De~Paris,  \altaffilmark{23}
% G.~Di~Cocco,  \altaffilmark{1}
% G.~Di~Persio,  \altaffilmark{4}
% A.~Ferrari,  \altaffilmark{14,15}
% M.~Fiorini,  \altaffilmark{7}
% T.~Froysland,  \altaffilmark{14,6}
% A.~Pellizzoni,  \altaffilmark{20}
% F.~Perotti,  \altaffilmark{7}
% P.~Picozza,  \altaffilmark{19}
% G.~Piano,   \altaffilmark{4,6,19}  % <-- added
% M.~Pilia,   \altaffilmark{16}  % <-- added
% M.~Prest,  \altaffilmark{16,17}
% G.~Pucella,  \altaffilmark{4}
% A.~Rubini,  \altaffilmark{4}
% S.~Sabatini,  \altaffilmark{19}
% E.~Striani,  \altaffilmark{4,6,19}  % <-- added
% E.~Vallazza,  \altaffilmark{5}
% V.~Vittorini,  \altaffilmark{4}
% A.~Zambra,  \altaffilmark{7,14}
% D.~Zanello,  \altaffilmark{10}
% L.A.~Antonelli,  \altaffilmark{9}
% D.~Gasparrini,  \altaffilmark{9}
% B.~Preger,  \altaffilmark{9}
% P.~Santolamazza,  \altaffilmark{9}
% L.~Salotti,  \altaffilmark{22}}

{1} {INAF-IASF, National Institute for Astrophysics, Bologna, Italy.}

{2} {Birkeland Centre for Space Science, Department of Physics and Technology, University of Bergen, Norway.}

{3} {INAF-IAPS Roma, via del Fosso del Cavaliere 100, I-00133 Roma, Italy.}

{4} {Dipartimento di Fisica, Universit\`a Tor Vergata, via della Ricerca Scientifica 1, I-00133 Roma, Italy.}

{5} {University of Agder, Department of Engineering Sciences, Norway.}

{6} {Dipartimento di Fisica e Astronomia, Universit\`a di Bologna, Bologna, Italy.}

{7} {ENEA, via Martiri di Monte Sole 4, I-40129 Bologna, Italy}

{8} {ASI Science Data Center, via del Politecnico snc, I-00133, Roma, Italy}

{9} {INAF-OAR, Via di Frascati, 33 I-00040, Monteporzio Catone (Roma), Italy}

{10} {Italian Space Agency, Via del Politecnico snc , 00133 Roma, Italy}

{11} {INAF-IASF Milano, via E. Bassini 15, I-20133 Milano, Italy}

{12} {INFN Pavia, via A. Bassi 6, I-27100 Pavia, Italy}

{13} {School of Physics, University of the Witwatersrand, Johannesburg Wits 2050, South Africa}

{14} {CNR-ISAC Roma, via del Fosso del Cavaliere 100, I-00133 Roma, Italy.}

% {13} {INAF, Viale del Parco Mellini 84, Roma, Italy}

{15} {Dipartimento di Fisica Universit\`a di Trieste, via A. Valerio 2, I-34127 Trieste, Italy}

{16} {INFN Trieste, via A. Valerio 2, I-34127 Trieste, Italy}

% {16} {INFN Roma ``Tor Vergata'', via della Ricerca Scientifica 1,}

% {17} {ENEA Frascati, via Enrico Fermi 45, I-00044 Frascati(Roma), Italy}

{17} {INAF-Osservatorio Astronomico di Cagliari, loc. Poggio dei Pini, strada 54, I-09012, Capoterra (CA), Italy}

% {19} {INAF-IASF Palermo, Via Ugo La Malfa 153, 90146 Palermo, Italy}

% {10} {INFN Roma ``La Sapienza'', p.le Aldo Moro 2, I-00185 Roma,
% Italy}

% {11} {Dipartimento di Fisica, Universit\`a La Sapienza, p.le Aldo
% Moro 2, I-00185 Roma, Italy}

% {14} {CIFS Torino, Viale Settimio Severo 63, I-10133 Torino, Italy}

% {15} {Dipartimento di Fisica, Universit\`a Torino, Torino, Italy}

% {16} {Dipartimento di Fisica, Universit\`a dell'Insubria, Via
% Valleggio 11, I-22100 Como, Italy}

% {17} {INFN Milano-Bicocca, Piazza della Scienza 3, I-20126 Milano,
% Italy}

% {18} {CNR-IMIP, Area della Ricerca di Montelibretti (Roma), Italy}

% I-00133 Roma, Italy}

% {21} {IKI, Moscow, Russia}

\clearpage
%
% \altaffiltext{2}{Department of Geography, Ohio State University,
% Columbus, Ohio, USA.}
%
% \altaffiltext{3}{Department of Space Sciences, University of Michigan,
% Ann Arbor, Michigan, USA.}
%
% \altaffiltext{4}{Desert Research Institute, Division of Hydrologic Sciences,
% Reno, Nevada, USA.}

%% ------------------------------------------------------------------------ %%
%
%  ABSTRACT
%
%% ------------------------------------------------------------------------ %%

% >> Do NOT include any \begin...\end commands within
% >> the body of the abstract.

\begin{abstract}

% 150 words
At the end of March 2015 the onboard software configuration of the AGILE satellite was modified in order to disable the veto signal of the anticoincidence shield for the minicalorimeter instrument. The motivation for such a change was the understanding that the dead time induced by the anticoincidence prevented the detection of a large fraction of Terrestrial Gamma-Ray Flashes (TGFs). The configuration change was highly successful resulting in an increase of one order of magnitude in TGF detection rate. As expected, the largest fraction of the new events has short duration ($< 100 \mathrm{\mu s}$), and part of them has simultaneous association with lightning sferics detected by the World Wide Lightning Location Network (WWLLN). The new configuration provides the largest TGF detection rate surface density (TGFs / $\mathrm{km^2}$ / year) to date, opening prospects for improved correlation studies with lightning and atmospheric parameters on short spatial and temporal scales along the equatorial region.

\end{abstract}

%% ------------------------------------------------------------------------ %%
%
%  BEGIN ARTICLE
%
%% ------------------------------------------------------------------------ %%

% The body of the article must start with a \begin{article} command
%
% \end{article} must follow the references section, before the figures
%  and tables.

\begin{article}
\section{Introduction}
\label{intro}

Terrestrial Gamma-ray Flashes (TGFs) are submillisecond bursts of
gamma-rays associated to lightning and thunderstorm activity and typically
observed from space. They represent the observable manifestation of
thunderstorm systems as the most energetic natural particle
accelerators on Earth \citep{Dwyer2012b}. Although many low-Earth orbiting
satellites equipped with gamma-ray detectors exist, TGF observations were routinely reported by only four of them: the Burst And Transient Source Experiment (BATSE) onboard the Compton Gamma-ray Observatory \citep{Fishman1994}, the Reuven Ramaty High-Energy Solar Spectroscopic Imager (RHESSI) \citep{Smith2005}, the Gamma-ray Burst Monitor (GBM) onboard the $Fermi$ Gamma-ray Space Telescope \citep{Briggs2010} and the Astrorivelatore Gamma a Immagini Leggero (AGILE) mission \citep{Marisaldi2010}, the last three currently operative. The reason for this relies on the typical time scale of this phenomenon ($\approx 100 \mathrm{\mu  s}$ average duration) which puts strong requirements on satellite data acquisition strategies and eventually on onboard trigger logic. In addition, the high TGF average fluence at satellite altitudes ($\approx 0.1 \mathrm{cm^{-2}}$ at 500--600~km) combined with their short duration makes all TGF detectors significantly affected by dead time and pile-up effects \citep{Grefenstette2009,Gjesteland2010,Briggs2010,Marisaldi2014}. These effects are instrument-dependent and must be carefully modeled and accounted for when trying to derive general unbiased properties of the observed TGF population.
 
\cite{Marisaldi2014} showed that AGILE TGF detections by the
minicalorimeter (MCAL) instrument were heavily affected by the dead
time induced by the anti-coincidence (AC) shield designed to reject
signals due to charged particles. Dead time in fact prevented the
detection of events with duration shorter than $\approx 100
\mathrm{\mu  s}$, biasing the duration distribution towards larger
values than observed by other spacecrafts. In addition, no precise
matches with radio signals located by the World Wide Lightning Location Network (WWLLN) were obtained, consistently with this chance being inversely proportional to the TGF duration \citep{Connaughton2013,Dwyer2013}. Moreover, since the AC is a paralyzable detector, the same set of observations can be due to events of different intrinsic fluence and duration, therefore no one-to-one dead time correction can be applied. As a result the TGF intensity distribution evaluated in \citep{Marisaldi2014}, distorted by dead time as well, was corrected by means of a forward folding approach based on assumptions on the intrinsic duration and fluence distributions.  

Based on the understanding that AC-induced dead time was significantly affecting the satellite TGF detection capabilities, the AGILE science team, in agreement with the Italian Space Agency (ASI) and in collaboration with the industrial partners and the ASI Science Data Center (ASDC), has undertaken the necessary steps to modify the onboard software configuration in order to inhibit the AC veto signal on the MCAL detector. Starting from 23 March 2015, the new configuration with AC veto disabled on MCAL has been steadily active onboard AGILE. The trigger logic parameters described in \cite{Marisaldi2014} were left unchanged except for the threshold on the 16~ms search time window, set from 22 to 41 counts to cope with the background rate increase. However, since all TGFs triggered on shorter time windows ($293 \mathrm{\mu  s}$ or 1~ms) this change does not affect the TGF trigger performance. In addition, the burst detection software is now not active during passage through the South Atlantic Anomaly (SAA). The following sections report the results on TGF detection after three months of operations in the new configuration. 

\section{Experimental results}

%------------------------------------------------------------

 \begin{figure}
 \noindent\includegraphics[width=3.5in]{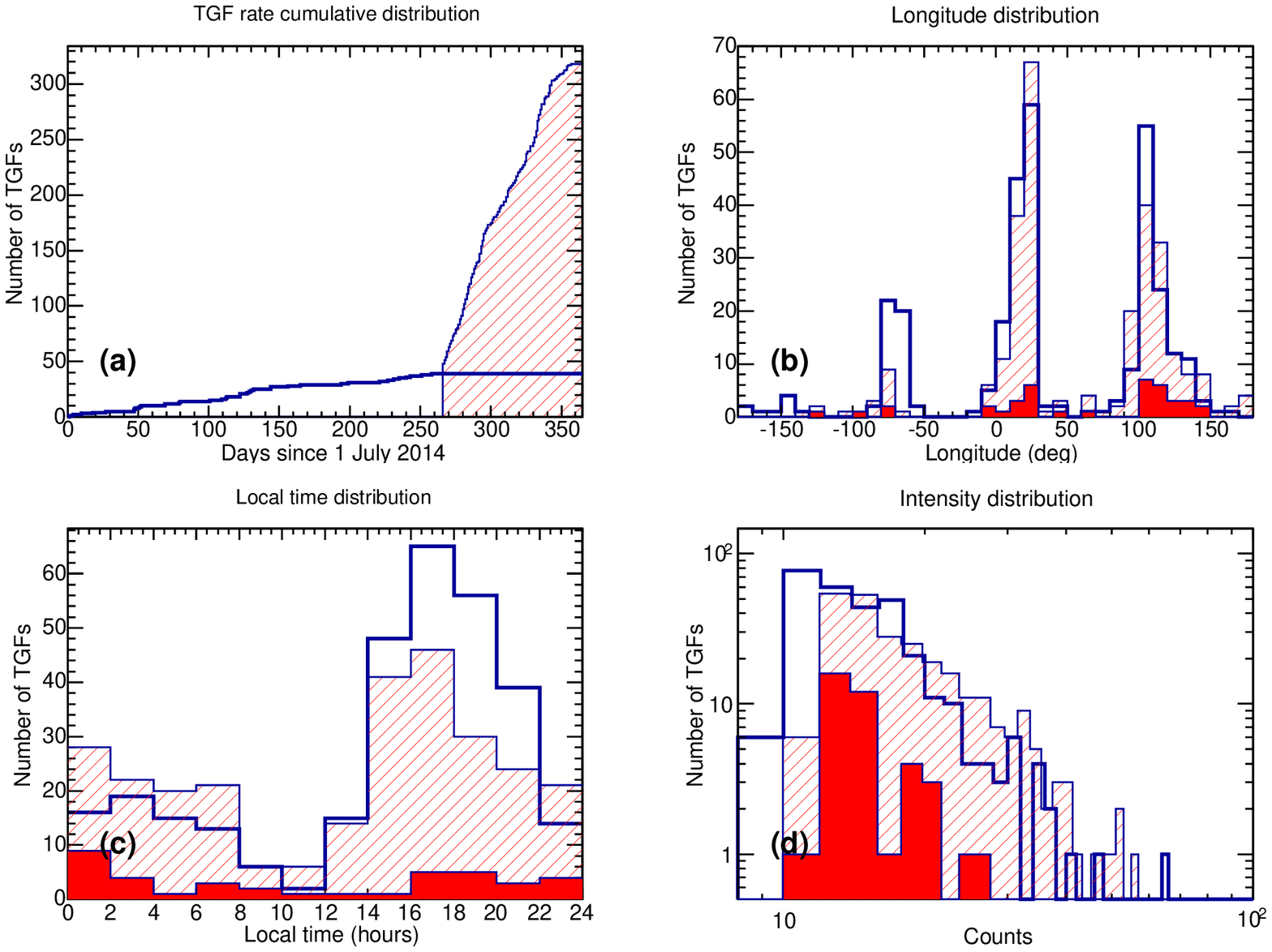}

\caption{Cumulative TGF rate (a), longitude (b), local time (c), number of counts (d) distributions for the TGF samples in standard (blue) and enhanced (red hatches) configurations. Red filled histograms are the corresponding distributions for events with a simultaneous WWLLN match.}
 \label{tgf_distr}
 \end{figure}

% created with ROOT macro: compare_tgf_newconf_PAPER.C

%------------------------------------------------------------

The same selection criteria on triggered data described in \cite{Marisaldi2014} were applied for TGF selection, thus allowing an unbiased comparison between the TGF samples obtained with the new (enhanced) and previous (standard) configurations. Between 23 March and 24 June 2015 a total of 279 TGFs have been recorded in the enhanced configuration. Figure \ref{tgf_distr} shows the cumulative detection rate and the distributions of longitude, local time and counts for the samples in both configurations.
The average daily TGF detection rate increased by one order of magnitude from 0.3 to 3 TGFs/day. Longitude and local time distributions clearly show the three continental lightning chimneys and the early morning / afternoon peaks, respectively, suggesting that the number of false events in the sample is low.
To support this statement we consider the ratio between the number of TGFs detected above a TGF-active region and a control region with low lightning activity and consequently expected low TGF detection rate, following the approach outlined in \cite{Briggs2013}. TGF-active region is defined as the three continental longitude bands (Central America: $[-90^{\circ} \, , \, -60^{\circ}]$; Africa: $[-10^{\circ} \, , \, +30^{\circ}]$; Maritime Continent: $[+100^{\circ} \, , \, +150^{\circ}]$). The control region is defined as the equatorial Pacific Ocean longitude band $[-140^{\circ} \, , \, -110^{\circ}]$ and has been chosen to be as close as possible to the Southeast Pacific control region defined in \cite{Briggs2013}.  Since in the case of AGILE the exposure on these regions is proportional to the longitude extent, we divide the number of observed TGFs in each region by this value. We obtain a TGF detection rate ratio of 20 between TGF-active and control regions, while \cite{Briggs2013} report a value 70 for the equivalent parameter. However, a direct comparison between these numbers cannot be done, mostly because of the different orbital inclination of the AGILE ($2.5^{\circ}$) and $Fermi$ ($25.6^{\circ}$) spacecrafts. In fact, while $Fermi$ control region well extends in the Southeast Pacific where lightning activity is very small \citep{Christian2003}, AGILE equatorial control region is close to the Intertropical Convergence Zone (ITCZ) were thunderstorm and lightning activity occur, especially during the observation period (March to May) \citep{Christian2003}.

%------------------------------------------------------------

 \begin{figure}
 \noindent\includegraphics[width=3.5in]{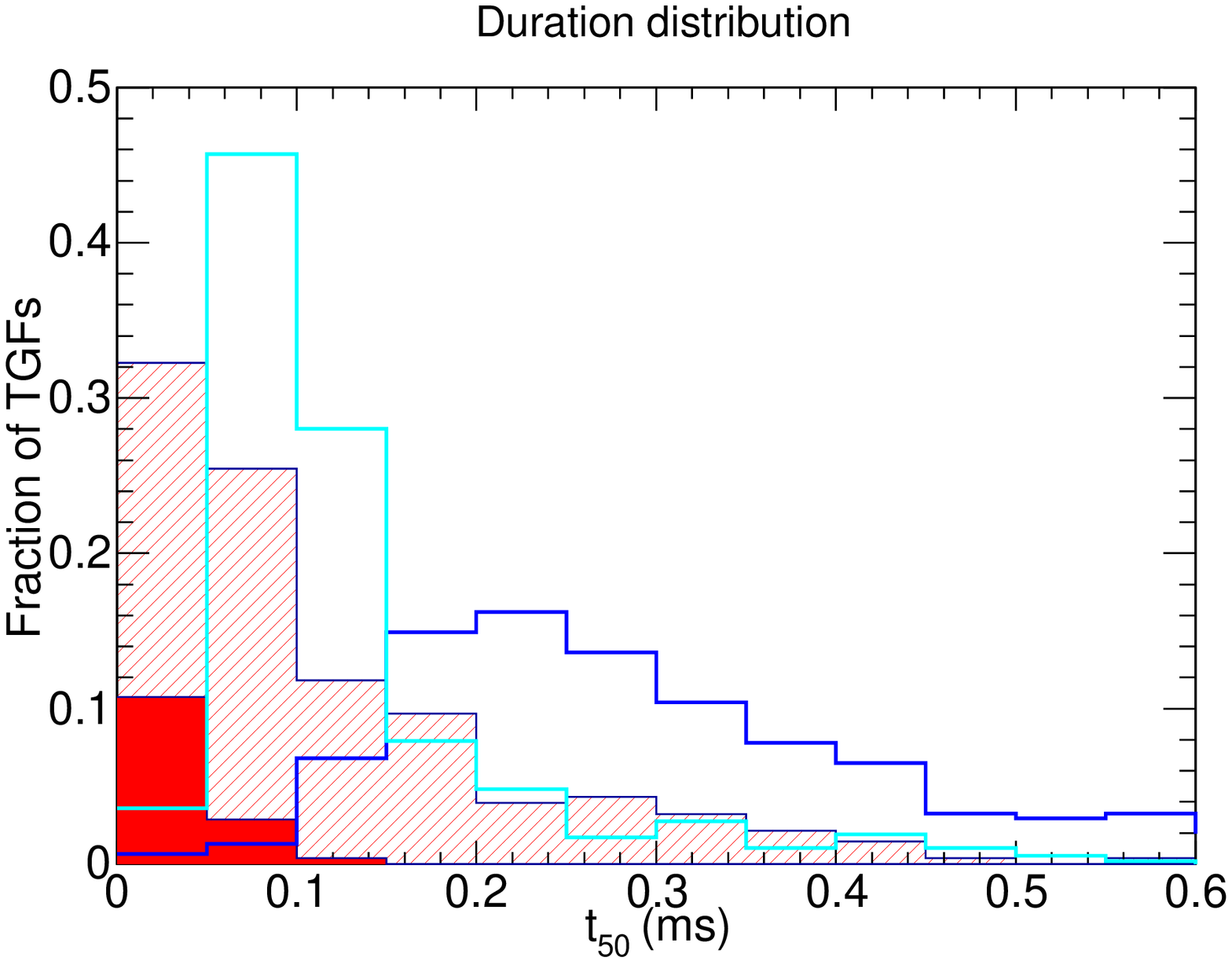}

\caption{Normalized duration ($t_{50}$) distribution for the enhanced (red hatches) and standard (blue line) TGF samples. Red filled histogram: $t_{50}$ distribution of the enhanced TGFs with a simultaneous WWLLN match normalized to the total number of the enhanced sample. Cyan histogram: $Fermi$ GBM $t_{50}$ distribution calculated for counts above 300~keV, from \cite{Connaughton2013} Fig.3.}
 \label{duration}
 \end{figure}

% created with ROOT macro: compare_tgf_newconf_PAPER.C

%------------------------------------------------------------

The time series of each TGF has been fitted by a Gaussian model superimposed to a constant background by means of the maximum likelihood technique in order to avoid the loss of information due to arbitrary time binning, as described in \cite{Marisaldi2014}. The duration and intensity of each TGF are then extracted by the model fit parameters. The TGF duration is calculated as $t_{50} = 1.349 \sigma$ and $t_{90} = 3.290 \sigma$, where  $t_{50}$ and  $t_{90}$ are defined as the central time intervals including the 50\% and 90\% of the counts, respectively, and $\sigma$ is the standard deviation of the Gaussian model. $t_{50}$ and  $t_{90}$ are convenient duration proxies when the paucity of counts does not allow a coherent identification of the start and end points of a transient.   
After close examination of the events light curves and fit results, we decided to exclude four events with poor convergence of the maximum likelihood procedure and seven events with closely spaced multiple peaks from the plots shown in Figure \ref{tgf_distr}d as well as in subsequent figures.
Figure \ref{duration} shows the duration distributions for the AGILE enhanced and standard samples and for $Fermi$ GBM events when only counts with energy larger than 300~keV are considered \citep{Connaughton2013}, to match the MCAL energy threshold. The duration distribution for TGFs with a close WWLLN match is also included, as discussed in the following. The median of the $t_{50}$ distribution is  $86 \mathrm{\mu s}$ and $290 \mathrm{\mu s}$ for the enhanced and standard configurations, respectively, clearly indicating that the enhanced configuration allows the detection of much shorter events than before, as expected.

%------------------------------------------------------------

 \begin{figure}
 \noindent\includegraphics[width=3.5in]{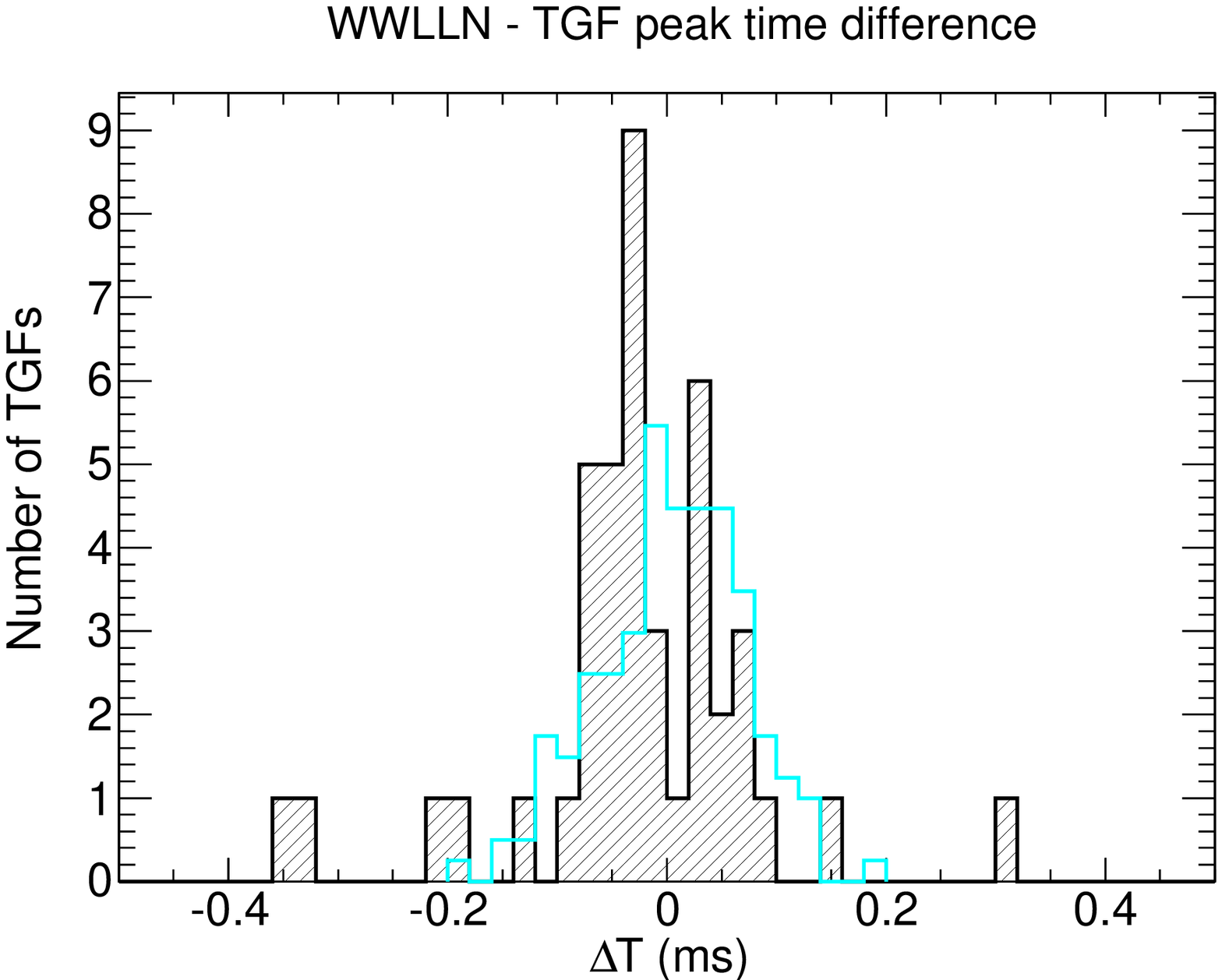}

\caption{Black hatches histogram: distribution of the time difference $\Delta T$ between the WWLLN detection closest in time to a TGF and the TGF peak time corrected for the light propagation time between the WWLLN location and the satellite.
%, assuming a production altitude of 15~km. A Gaussian fit to the distribution is included. 
Cyan line: same distribution for $Fermi$ events as published in \cite{Connaughton2013}, Fig. 1, normalized to the number of AGILE events for visualization purposes.}
 \label{wwlln}
 \end{figure}

% created with ROOT macro: compare_tgf_newconf_PAPER.C

%------------------------------------------------------------

The database of lightning detected by the World Wide Lightning
Location Network (WWLLN) \citep{Rodger2009} has been searched for
correlation with the enhanced TGF sample. Figure \ref{wwlln} shows the
distribution of the time difference $\Delta T$ between the WWLLN
detection closest in time to a TGF and the TGF peak time corrected for
the light propagation time between the WWLLN location and the
satellite, assuming a source production altitude of 15~km \citep{Dwyer2005}. 
Given the low number of counts, the peak time of a TGF is defined by the centroid of the
Gaussian model fit described above. A total number of 39 sferics
within $200 \mathrm{\mu s}$ from the TGF peak time has been observed, hereafter defined as simultaneous sferics according to the definition in \cite{Connaughton2013},
corresponding to 14\% of the sample. 
%The distribution can be fit with a Gaussian with centroid $\Delta T_0 = -35 \mathrm{\mu s}$ and standard deviation $\sigma = 28 \mathrm{\mu s}$. 
The same histogram obtained for $Fermi$ data and shown in Figure~1 of \cite{Connaughton2013}, normalized to the number of AGILE events, is also shown for reference.

\section{Discussion}
\label{discussion}

The inhibition of the AC veto for the AGILE MCAL instrument has clearly resulted in the detection of a much larger number of TGFs than with the standard configuration, enhancing the sensitivity for events lasting less than $100 \mathrm{\mu s}$. These results validate the data interpretation provided in \cite{Marisaldi2014} regarding the role of dead time in biasing the observed duration and intensity distributions and the lack of detection of simultaneous sferics.

%------------------------------------------------------------

 \begin{figure}
 \noindent\includegraphics[width=3.5in]{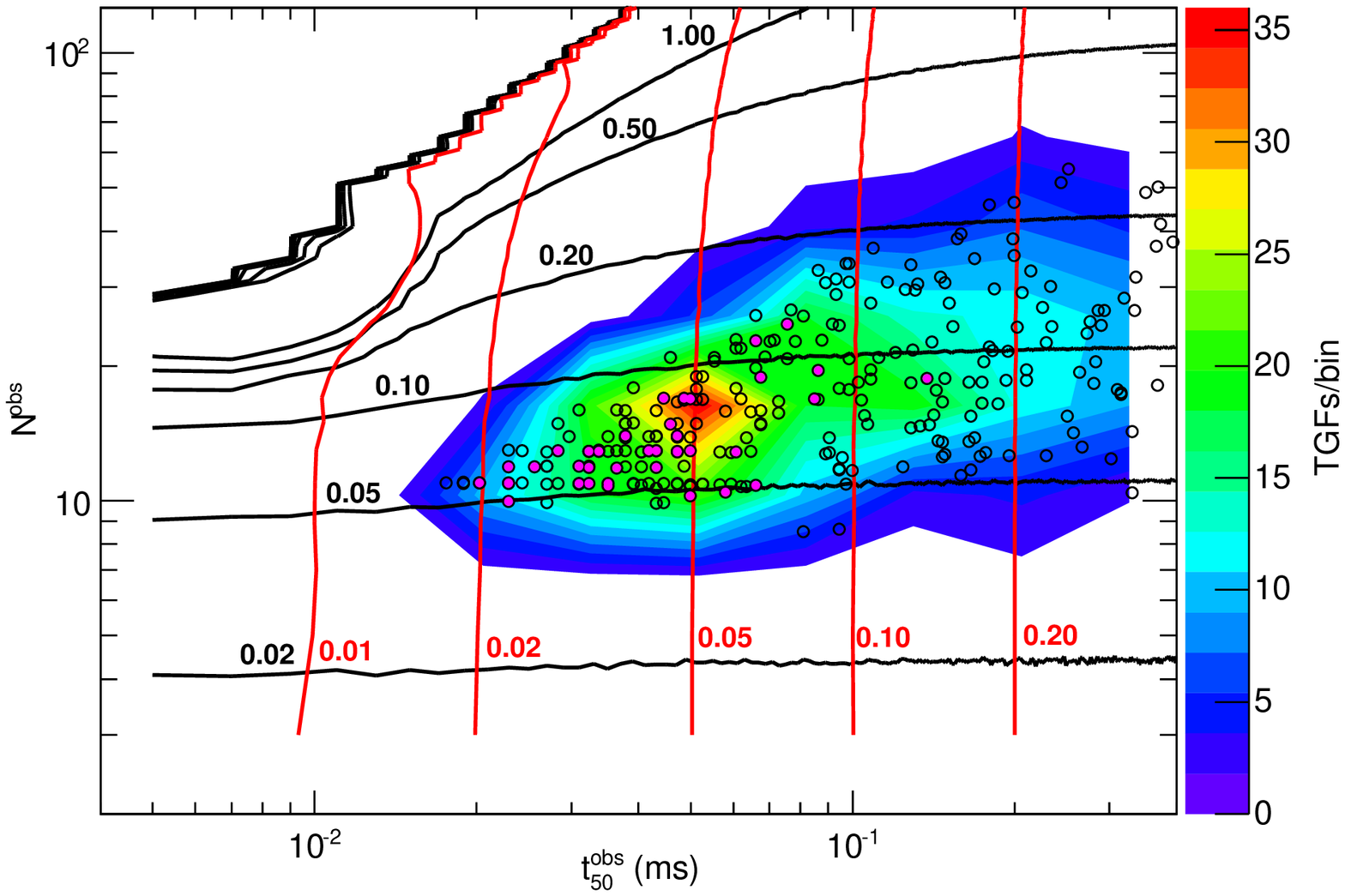}

\caption{Enhanced sample (black circles) in the observed $(N^{obs}
  \times t_{50}^{obs})$, i.e. counts $\times$ duration, parameter
  space. The color contour highlights the region with the highest
  detection rate (five logarithmic bins per decade in both coordinates). 
Magenta dots are the events with a simultaneous WWLLN match.
Red and black contours mark points corresponding to the same value of $t_{50}$ (ms) and fluence $F$ ($cm^{-2}$), respectively, according to the MCAL dead time model for $60^\circ$ off-axis angle. To obtain $N^{obs}$, $F$ must be multiplied times the effective area of $\mathrm{\approx 220\, cm^2}$.}
 \label{deadtime}
 \end{figure}

% created with ROOT macro: compare_tgf_newconf_PAPER.C

%------------------------------------------------------------

Once the contribution by the AC has been removed, the major source of dead time in the enhanced configuration is the data readout of the scintillation bars comprising MCAL. Each of the 30 bars (26 were active at the time the data presented here were collected) acts as an independent nonparalyzable detector, requiring $20\, \mathrm{\mu s}$ for a photon acquisition after the discriminator has fired. Any detector trigger occuring during this fixed time is rejected. 
Although the dead time per count is larger than that reported for $Fermi$ GBM ($2.6\, \mathrm{\mu s}$, \citep{Briggs2013}), the key point to overall dead time effects mitigation is the MCAL spatial segmentation. Since each of the MCAL bars is independent from the others, two consecutive photons separated in time less than $20\, \mathrm{\mu s}$ are promptly recorded, provided they hit two different bars ready for data acquisition.
We addressed the issue of MCAL-induced dead time by means of dedicated Monte Carlo simulations, described in details in \cite{Marisaldi2014}, using the full AGILE mass model and a typical TGF spectrum as reported in \cite{Dwyer2005}. 
We considered a TGF as described by two observables: its fluence at satellite altitude $F$ and its duration expressed in terms of $t_{50}$. 
We then consider the combined effect of detector effective area ($\mathrm{A_{eff} \approx 220\, cm^2}$ for a typical TGF spectrum incident at $60^\circ$ off-axis angle) and dead time as a function $f$ that maps the $(F \, \times \, t_{50})$ into the $(N^{obs} \, \times \, t_{50}^{obs})$ space, where $N^{obs}$ and $t_{50}^{obs}$ are the measured number of counts and duration.
The effective area for $60^\circ$ off-axis angle has been chosen as an average value for all the possible incoming off-axis angles. The maximum effective area, corresponding to the photon beam hitting orthogonal to the MCAL plane is just 15\% larger. 
We note that, since the detectors are nonparalyzable, the function $f$ can be inverted to extract the expected true fluence and duration from the observed counts and duration, i.e. each TGF can be individually corrected for dead time. This situation is radically different from  the standard configuration with the AC acting as a paralyzable detector.
In order to build the function $f$ we generated $10^7$ simulated TGFs uniformly distributed in the parameter space given by $ (0.01\, \mathrm{cm^{-2}} < F < 4\, \mathrm{cm^{-2}}) \times  (0.002\, \mathrm{ms} < t_{50} < 0.4\, \mathrm{ms})$. For each simulated TGF the expected number of counts in MCAL was defined according to $F$ and the average effective area given by simulation results; the time series of the counts were randomly extracted according to a Gaussian time profile with $\sigma = 0.74\, t_{50}$; and finally each count was assigned to a detector bar based on a uniform random distribution. We then apply to the MCAL time series the effect of the dead time induced by the MCAL detector processing time, rejecting all counts within a $20\, \mu$s time interval following a count on the same scintillating bar. For each simulated TGF we then count the observed number of counts $N^{obs}$ and evaluate the observed duration $t_{50}^{obs}$ by calculating the standard deviation of the counts time series, multiplied by factor 1.349 to convert from standard deviation to $t_{50}$.
With respect to the simulations described in \cite{Marisaldi2014}, we modified the code to account for a 50\% fraction of double counts, i.e. counts involving two bars, mostly due to Compton scattering of photons between neighboring bars, which affects the number of active bars and can eventually increase the dead time fraction. The chosen double counts fraction is a worst case estimate, the typical fraction for TGFs being of order of $\approx 30\%$. Although this change, we do not observe a significant variation with respect to previous simulations.
Figure \ref{deadtime} presents the enhanced TGF sample in the $(N^{obs} \, \times \, t_{50}^{obs})$ parameter space. The lines corresponding to true fluence and duration values are superimposed. As a rule of thumb, dead time is significantly affecting the sample in those regions where the lines deviate significantly from a parallel grid. 
The effect of dead time is that the observed number of photons in a TGF is less than the true number of photons that hit the detector. Also the estimated duration of the TGF gets longer as the losses due to dead time increase. With the current settings the fluence at detection threshold is $\approx 0.05\, \mathrm{cm^{-2}}$ as evidenced by the TGF population shown in Figure \ref{deadtime}.

The median of the $t_{50}$ distribution for the enhanced sample is $86\, \mathrm{\mu s}$, significantly shorter than the   $290\, \mathrm{\mu s}$ obtained for the standard sample. This median value is also shorter than the $100\, \mathrm{\mu s}$ reported for the $Fermi$ GBM sample \citep{Briggs2013}, but a greater evidence for the difference in the duration distribution for the two samples is obtained if we consider the fraction of events with $t_{50} \le 50 \mathrm{\mu s}$, as shown in Figure \ref{duration}. \cite{Briggs2013} noted the lack of very short events in the $Fermi$ sample suggesting it was a consequence of dead time. According to our dead time model the observed AGILE enhanced sample is in a region where the effect of dead time is very low and substantially negligible with respect to the errors due to counting statistics, as shown in Figure \ref{deadtime}. 
We note from Figure \ref{deadtime} that high-fluence TGFs tend to have a longer duration than dimmer ones, e.g. no TGF with $F \ge 0.1\, \mathrm{cm^{-2}}$ is found with $ t_{50} \le 50\, \mathrm{\mu s}$. According to our dead time model, this is not an observational bias due to dead time, but a physical feature of the TGF population that could be a test bench for production models. For example, \cite{Dwyer2012} reports that, in the Relativistic Feedback (RF) model longer TGFs produce less runaway electrons than shorter ones, for a given electric field configuration. This is apparently in contrast with our observations. Conversely, the hypothesis that longer TGFs are the result of closely spaced shorter events can be ruled out based on the work by \cite{Connaughton2013}, where an inverse proportionality between TGF duration and the likelihood of association with WWLLN sferics is firmly established. 
Whereas the detection of low fluence events is limited by the detector effective area, there is no instrumental issue preventing the detection of very short events with duration of $\approx 20\, \mathrm{\mu s}$ or less. 
However, we cannot neglect the contribution of Compton scattering of photons through the atmosphere to the observed duration of TGFs at satellite altitude \citep{Ostgaard2008,Grefenstette2008,Celestin2012b,Fitzpatrick2014}. Although this effect is more evident for low energy photons below MCAL threshold, \cite{Celestin2012b} showed that also instantaneously produced 1~MeV photons can be spread due to Compton scattering over a time interval as long as $50\, \mathrm{\mu s}$ at satellite altitude. Therefore, the observed $\approx 20\, \mathrm{\mu s}$ lower limit in TGF duration may be a measure of the minimum incompressible time spread due to Compton scattering rather than the intrinsic lower limit of TGF duration.

Concerning the correlation with WWLLN sferics, we first note that no simultaneous sferics was observed for the standard sample \citep{Marisaldi2014} and the reason for that was correctly identified in the bias towards longer events due to dead time suppression of short events, the latter of which are more likely associated to lightning sferics according to \cite{Connaughton2013}. 
The characteristics of the sample with simultaneous WWLLN sferics follow the general pattern discussed in \cite{Connaughton2013}, with few significant quantitative differences. The overall fraction of TGFs with simultaneous sferics is 14\% while it is 30\% in the $Fermi$ sample. 
This fraction rises to 33\% for TGF duration shorter than $50\, \mathrm{\mu s}$, to be compared to the 60\% value reported by \cite{Connaughton2013}.
The anticorrelation between TGF duration and the probability of being associated to a simultaneous sferics is confirmed, as can be seen in Figure \ref{duration}. 
The difference in overall simultaneous sferics detection probability may reside in the different orbital inclination of the AGILE and $Fermi$ missions, that make the two spacecrafts cover regions where the WWLLN detection efficiency is remarkably different. In particular, the WWLLN detection efficiency over equatorial Africa, where $\approx 40\%$ of the AGILE TGFs are observed, can be on average the 20\% of the efficiency over North America and the Pacific region \citep{Hutchins2012}, i.e. about 25\% of the efficiency for the other low latitude regions. This scenario is supported by the low number of simultaneous detections observed over Africa, as shown in Figure \ref{tgf_distr}. 
If we assume the WWLLN detection efficiency on the remaining equatorial regions covered by AGILE orbit is $\approx 80\%$ the average value for $Fermi$, which is reasonable considering the plots in \cite{Hutchins2012}, we can estimate the ratio between the number of TGFs with WWLLN simultaneous match for AGILE and $Fermi$ to be $R_{AF} = (0.4 \times 0.25 + 0.6) \times 0.8 = 0.56$. Although very simplified, this estimate is remarkably consistent with the observed value $0.33 / 0.60 = 0.55$ obtained for TGF with duration shorter than $50\, \mathrm{\mu s}$.
However, WWLLN efficiency significantly changes as a function of local time and generally improves over time as long as new stations are added to the network, therefore a more detailed comparison between AGILE and $Fermi$ association rate is difficult. 

The WWLLN-TGF time difference $\mathrm{\Delta t}$ distribution shown in Figure \ref{wwlln} {could appear to be bimodal, although} we found no correlation between $\mathrm{\Delta t}$ and any TGF parameter. We therefore regard this feature as due to the limited statistics. The root mean square (RMS) of the $\mathrm{\Delta t}$ distribution within $\pm 200\, \mathrm{\mu s}$ is $63\, \mathrm{\mu s}$. This error value can be regarded as the quadrature sum of several independent timing error contributions, namely the WWLLN accuracy ($\approx 15\, \mathrm{\mu s}$ \citep{Hutchins2012}), the error on the TGF peak determination by the maximum likelihood procedure ($\approx 10\, \mathrm{\mu s}$) and the uncertainty on TGF production height ($\sim 10\, \mathrm{km}\, \approx 30\, \mathrm{\mu s}$). The uncertainty on the GPS-provided AGILE position and the MCAL timing accuracy ($\approx 2\, \mathrm{\mu s}$) add negligible contributions. An additional timing uncertainty term of order of $\approx 50\, \mathrm{\mu s}$ is still missing to justify the observed $\mathrm{\Delta t}$ distribution RMS, if we assume that the TGF-producing electrons are responsible also for the sferics radio pulses \citep{Connaughton2013,Dwyer2013}. 
However \cite{Cummer2011} showed that the electromagnetic waveform associated to TGFs can be very complex, containing several fast pulses, and that the pulse corresponding to the localization by ground-based lightning location networks may not be the closest in time to the TGF. In fact, this may be the source of the missing term in the uncertainty analysis discussed above. Therefore, even if we can narrow down to $\pm 100\, \mathrm{\mu s}$ the definition for simultaneity, provided the distribution of Figure \ref{wwlln} is confirmed with higher statistics, it will be difficult to further improve this value to gather more information on the relative timing between lightning and TGF.
% from these data unless uncertainties on lightning location and TGF production altitude are reduced.

\section{Conclusion}

The enhanced configuration, which has been running onboard AGILE since 23 March 2015, has been highly successful resulting in a TGF detection rate increase of about one order of magnitude and opening up unique scientific opportunities for the understanding of the relationship between TGFs and lightning activity. For the first time, we present a TGF sample unbiased by dead time effects, which may serve as a test bench for production models. We also provide an independent confirmation of the anticorrelation between TGF duration and the likelihood of association with simultaneous WWLLN lightning sferics reported in \citep{Connaughton2013}.
The large number of expected events, $\approx 1000$/year, larger than that of $Fermi$ GBM in continuous Time-Tagged Event data acquisition mode \citep{Briggs2013}, concentrated in a narrow latitude band across the equator thanks to the AGILE orbital inclination of $2.5^{\circ}$, provides an unprecedently high TGF detection rate surface density, i.e. number of TGFs observed per unit area per unit time. In turn, this should allow enough counting statistics for correlation studies with lightning activity on small spatial and time scales. Up to now, only correlation studies on timescales of several years have been reported \citep{Smith2010,Fuschino2011} so any seasonal variability giving possible clues on the TGF/lightning relation has been smeared out. 

We recall here that the selection criteria used in this work for TGF identification are the same as those previously used in \cite{Marisaldi2014}, including the requirement for the maximum photon energy not to be greater than 30~MeV. The availability of a sample of events with a firm association to WWLLN sferics, therefore surely associated to a TGF process, will allow to relax all selection criteria including the cut on the maximum photon energy, possibly shedding light on the long-standing issue of the highest photon energy in TGFs \citep{Tavani2011}.

%------------------------------------------------------------

\appendix

%%% End of body of article:

%%%%%%%%%%%%%%%%%%%%%%%%%%%%%%%%
%% Optional Appendix goes here
%
%%%%%%%%%%%%%%%%%
% Geophysical Research Letters only allows an appendix without a letter.
%% You can get this result with 
%  \section*{Appendix} 
%  or 
%  \section*{Appendix: Title}
%%%%%%%%%%%%%%%%%
%
% \appendix resets counters and redefines section heads
% but doesn't print anything. 
% After typing  \appendix 
%
% \section{Here Is Appendix Title}
% will print 
% Appendix A: Here Is Appendix Title
%
% \section*{Appendix}
% will print 
% Appendix 
%
% \section*{Appendix: Here Is Appendix Title}
% will print 
% Appendix: Here Is Appendix Title 
%
% For only 1 appendix \appendix \section{Appendix} is preferred.
% which will print 
% Appendix A

%%%%%%%%%%%%%%%%%%%%%%%%%%%%%%%%%%%%%%%%%%%%%%%%%%%%%%%%%%%%%%%%
%
% Optional Glossary or Notation section, goes here
%
%%%%%%%%%%%%%%
% Glossary only allowed in Reviews of Geophysics
% \section*{Glossary}
% \paragraph{Term}
% Term Definition here
%
%%%%%%%%%%%%%%
% Notation -- End each entry with a period.
% \begin{notation}
% Term & definition.\\
% Second Term & second definition.
% \end{notation}
%%%%%%%%%%%%%%%%%%%%%%%%%%%%%%%%%%%%%%%%%%%%%%%%%%%%%%%%%%%%%%%%
%
%  ACKNOWLEDGMENTS

\begin{acknowledgments}
AGILE is a mission of the Italian Space Agency (ASI), with co-participation
of INAF (Istituto Nazionale di Astrofisica) and INFN (Istituto
Nazionale di Fisica Nucleare). This work was carried out in the frame
of the ASI-INAF agreement I/028/12/0.
This study was supported by the European Research Council under the European Union's Seventh Framework
Programme (FP7/2007-2013)/ERC grant agreement n. 320839 and the Research Council of Norway under 
contracts 208028/F50, 216872/F50 and 223252/F50 (CoE).
The authors wish to thank the World Wide Lightning Location Network (http://wwlln.net), a collaboration among over 50 universities and institutions, for providing the lightning location data used in this paper.
The properties of the TGF sample presented in this work are publicly available at the ASI Science Data Center (ASDC) website: http://www.asdc.asi.it/mcaltgfcat/ All other MCAL data used in this work are available upon request from M. Marisaldi (marisaldi@iasfbo.inaf.it).
The authors thank the AGILE industrial team at Compagnia Generale per lo Spazio (CGS) and Telespazio for their support during the configuration change. 
The authors also thank Valerie Connaughton for helpful support.
This work is dedicated to the memory of Paolo Sabatini, AGILE Program Manager at CGS, whose effort largely contributed to the success of the AGILE mission.
\end{acknowledgments}

%% ------------------------------------------------------------------------ %%
%
%  REFERENCE LIST AND TEXT CITATIONS
%
% Either type in your references using
% \begin{thebibliography}{}
% \bibitem
% Text
% \end{thebibliography}
%
% Or, 
%
% If you use BiBTeX for your References, please produce your .bbl
% file and copy the contents into your paper here.
%
% Follow these steps:
% 1. Run LaTeX on your LaTeX file.
%
% 2. Run BiBTeX on your LaTeX file.
%
% 3. Open the new .bbl file containing the reference list and
%   copy all the contents into your LaTeX file here.
%
% 4. Comment out the old \bibliographystyle and \bibliography commands.
%
% 5. Run LaTeX on your new file before submitting.
%
% AGU does not want a .bib or a .bbl file, but asks that you
% copy in the contents of your .bbl file here.

%\bibliographystyle{agu08}
%\bibliography{bibliography_MM_3}

\begin{thebibliography}{26}
\providecommand{\natexlab}[1]{#1}
\expandafter\ifx\csname urlstyle\endcsname\relax
  \providecommand{\doi}[1]{doi:\discretionary{}{}{}#1}\else
  \providecommand{\doi}{doi:\discretionary{}{}{}\begingroup
  \urlstyle{rm}\Url}\fi

\bibitem[{\textit{{Briggs} et~al.}(2010)}]{Briggs2010}
{Briggs}, M.~S., et~al. (2010), {First results on terrestrial gamma ray flashes
  from the Fermi Gamma-ray Burst Monitor}, \textit{Journal of Geophysical
  Research (Space Physics)}, \textit{115}, A07323, \doi{10.1029/2009JA015242}.

\bibitem[{\textit{{Briggs} et~al.}(2013)}]{Briggs2013}
{Briggs}, M.~S., et~al. (2013), {Terrestrial gamma-ray flashes in the Fermi
  era: Improved observations and analysis methods}, \textit{Journal of
  Geophysical Research (Space Physics)}, \textit{118}, 3805--3830,
  \doi{10.1002/jgra.50205}.

\bibitem[{\textit{{Celestin} and {Pasko}}(2012)}]{Celestin2012b}
{Celestin}, S., and V.~P. {Pasko} (2012), {Compton scattering effects on the
  duration of terrestrial gamma-ray flashes}, \textit{\grl}, \textit{39},
  L02802, \doi{10.1029/2011GL050342}.

\bibitem[{\textit{{Christian} et~al.}(2003)}]{Christian2003}
{Christian}, H.~J., et~al. (2003), {Global frequency and distribution of
  lightning as observed from space by the Optical Transient Detector},
  \textit{Journal of Geophysical Research (Atmospheres)}, \textit{108}, 4005,
  \doi{10.1029/2002JD002347}.

\bibitem[{\textit{{Connaughton} et~al.}(2013)}]{Connaughton2013}
{Connaughton}, V., et~al. (2013), {Radio signals from electron beams in
  terrestrial gamma ray flashes}, \textit{Journal of Geophysical Research
  (Space Physics)}, \textit{118}, 2313--2320, \doi{10.1029/2012JA018288}.

\bibitem[{\textit{{Cummer} et~al.}(2011)\textit{{Cummer}, {Lu}, {Briggs},
  {Connaughton}, {Xiong}, {Fishman}, and {Dwyer}}}]{Cummer2011}
{Cummer}, S.~A., G.~{Lu}, M.~S. {Briggs}, V.~{Connaughton}, S.~{Xiong}, G.~J.
  {Fishman}, and J.~R. {Dwyer} (2011), {The lightning-TGF relationship on
  microsecond timescales}, \textit{\grl}, \textit{38}, L14810,
  \doi{10.1029/2011GL048099}.

\bibitem[{\textit{{Dwyer}}(2012)}]{Dwyer2012}
{Dwyer}, J.~R. (2012), {The relativistic feedback discharge model of
  terrestrial gamma ray flashes}, \textit{Journal of Geophysical Research
  (Space Physics)}, \textit{117}, A02308, \doi{10.1029/2011JA017160}.

\bibitem[{\textit{{Dwyer} and {Cummer}}(2013)}]{Dwyer2013}
{Dwyer}, J.~R., and S.~A. {Cummer} (2013), {Radio emissions from terrestrial
  gamma-ray flashes}, \textit{Journal of Geophysical Research (Space Physics)},
  \textit{118}, 3769--3790, \doi{10.1002/jgra.50188}.

\bibitem[{\textit{{Dwyer} and {Smith}}(2005)}]{Dwyer2005}
{Dwyer}, J.~R., and D.~M. {Smith} (2005), {A comparison between Monte Carlo
  simulations of runaway breakdown and terrestrial gamma-ray flash
  observations}, \textit{Geophys.~Res.~Lett.}, \textit{32}, 22,804--+,
  \doi{10.1029/2005GL023848}.

\bibitem[{\textit{{Dwyer} et~al.}(2012)\textit{{Dwyer}, {Smith}, and
  {Cummer}}}]{Dwyer2012b}
{Dwyer}, J.~R., D.~M. {Smith}, and S.~A. {Cummer} (2012), {High-Energy
  Atmospheric Physics: Terrestrial Gamma-Ray Flashes and Related Phenomena},
  \textit{Space Science Reviews}, \textit{173}, 133--196,
  \doi{10.1007/s11214-012-9894-0}.

\bibitem[{\textit{Fishman et~al.}(1994)}]{Fishman1994}
Fishman, G.~J., et~al. (1994), {Discovery of intense gamma-ray flashes of
  atmospheric origin}, \textit{Science}, \textit{264}, 1313--1316.

\bibitem[{\textit{{Fitzpatrick} et~al.}(2014)}]{Fitzpatrick2014}
{Fitzpatrick}, G., et~al. (2014), {Compton scattering in terrestrial gamma-ray
  flashes detected with the Fermi gamma-ray burst monitor}, \textit{Physical
  Review D}, \textit{90}(4), 043008, \doi{10.1103/PhysRevD.90.043008}.

\bibitem[{\textit{{Fuschino} et~al.}(2008)}]{Fuschino2008}
{Fuschino}, F., et~al. (2008), {Search of GRB with AGILE Minicalorimeter},
  \textit{Nuclear Instruments and Methods in Physics Research A}, \textit{588},
  17--21, \doi{10.1016/j.nima.2008.01.004}.

\bibitem[{\textit{{Fuschino} et~al.}(2011)}]{Fuschino2011}
{Fuschino}, F., et~al. (2011), {High spatial resolution correlation of AGILE
  TGFs and global lightning activity above the equatorial belt}, \textit{\grl},
  \textit{38}, L14806, \doi{10.1029/2011GL047817}.

\bibitem[{\textit{{Gjesteland} et~al.}(2010)\textit{{Gjesteland},
  {{\O}stgaard}, {Connell}, {Stadsnes}, and {Fishman}}}]{Gjesteland2010}
{Gjesteland}, T., N.~{{\O}stgaard}, P.~H. {Connell}, J.~{Stadsnes}, and G.~J.
  {Fishman} (2010), {Effects of dead time losses on terrestrial gamma ray flash
  measurements with the Burst and Transient Source Experiment}, \textit{Journal
  of Geophysical Research (Space Physics)}, \textit{115}, A00E21,
  \doi{10.1029/2009JA014578}.

\bibitem[{\textit{{Grefenstette} et~al.}(2008)\textit{{Grefenstette}, {Smith},
  {Dwyer}, and {Fishman}}}]{Grefenstette2008}
{Grefenstette}, B.~W., D.~M. {Smith}, J.~R. {Dwyer}, and G.~J. {Fishman}
  (2008), {Time evolution of terrestrial gamma ray flashes},
  \textit{Geophys.~Res.~Lett.}, \textit{35}, 6802--+,
  \doi{10.1029/2007GL032922}.

\bibitem[{\textit{{Grefenstette} et~al.}(2009)\textit{{Grefenstette}, {Smith},
  {Hazelton}, and {Lopez}}}]{Grefenstette2009}
{Grefenstette}, B.~W., D.~M. {Smith}, B.~J. {Hazelton}, and L.~I. {Lopez}
  (2009), {First RHESSI terrestrial gamma ray flash catalog}, \textit{J.
  Geophys. Res.}, \textit{114}, A02314, \doi{10.1029/2008JA013721}.

\bibitem[{\textit{{Hutchins} et~al.}(2012)\textit{{Hutchins}, {Holzworth},
  {Brundell}, and {Rodger}}}]{Hutchins2012}
{Hutchins}, M.~L., R.~H. {Holzworth}, J.~B. {Brundell}, and C.~J. {Rodger}
  (2012), {Relative detection efficiency of the World Wide Lightning Location
  Network}, \textit{Radio Science}, \textit{47}, RS6005,
  \doi{10.1029/2012RS005049}.

\bibitem[{\textit{{Labanti} et~al.}(2009)}]{Labanti2009}
{Labanti}, C., et~al. (2009), {Design and construction of the Mini-Calorimeter
  of the AGILE satellite}, \textit{Nuclear Instruments and Methods in Physics
  Research A}, \textit{598}, 470--479, \doi{10.1016/j.nima.2008.09.021}.

\bibitem[{\textit{{Marisaldi} et~al.}(2010)}]{Marisaldi2010}
{Marisaldi}, M., et~al. (2010), {Detection of terrestrial gamma ray flashes up
  to 40 MeV by the AGILE satellite}, \textit{Journal of Geophysical Research
  (Space Physics)}, \textit{115}, A00E13, \doi{10.1029/2009JA014502}.

\bibitem[{\textit{Marisaldi et~al.}(2014)}]{Marisaldi2014}
Marisaldi, M., et~al. (2014), {Properties of terrestrial gamma ray flashes
  detected by AGILE MCAL below 30 MeV}, \textit{Journal of Geophysical
  Research: Space Physics}, \textit{119}(2), 1337--1355,
  \doi{10.1002/2013JA019301}.

\bibitem[{\textit{{{\O}stgaard} et~al.}(2008)\textit{{{\O}stgaard},
  {Gjesteland}, {Stadsnes}, {Connell}, and {Carlson}}}]{Ostgaard2008}
{{\O}stgaard}, N., T.~{Gjesteland}, J.~{Stadsnes}, P.~H. {Connell}, and
  B.~{Carlson} (2008), {Production altitude and time delays of the terrestrial
  gamma flashes: Revisiting the Burst and Transient Source Experiment spectra},
  \textit{J. Geophys. Res.}, \textit{113}, A02307, \doi{10.1029/2007JA012618}.

\bibitem[{\textit{{Rodger} et~al.}(2009)\textit{{Rodger}, {Brundell},
  {Holzworth}, and {Lay}}}]{Rodger2009}
{Rodger}, C.~J., J.~B. {Brundell}, R.~H. {Holzworth}, and E.~H. {Lay} (2009),
  {Growing Detection Efficiency of the World Wide Lightning Location Network},
  in \textit{American Institute of Physics Conference Series}, \textit{American
  Institute of Physics Conference Series}, vol. 1118, pp. 15--20,
  \doi{10.1063/1.3137706}.

\bibitem[{\textit{{Smith} et~al.}(2010)\textit{{Smith}, {Hazelton},
  {Grefenstette}, {Dwyer}, {Holzworth}, and {Lay}}}]{Smith2010}
{Smith}, D.~M., B.~J. {Hazelton}, B.~W. {Grefenstette}, J.~R. {Dwyer}, R.~H.
  {Holzworth}, and E.~H. {Lay} (2010), {Terrestrial gamma ray flashes
  correlated to storm phase and tropopause height}, \textit{Journal of
  Geophysical Research (Space Physics)}, \textit{115}, A00E49,
  \doi{10.1029/2009JA014853}.

\bibitem[{\textit{Smith et~al.}(2005)}]{Smith2005}
Smith, D.~M., et~al. (2005), {Terrestrial gamma-ray flashes observed up to 20
  MeV}, \textit{Science}, \textit{307}, 1085--1088.

\bibitem[{\textit{{Tavani} et~al.}(2011)}]{Tavani2011}
{Tavani}, M., et~al. (2011), {Terrestrial Gamma-Ray Flashes as Powerful
  Particle Accelerators}, \textit{Physical Review Letters}, \textit{106}(1),
  018501, \doi{10.1103/PhysRevLett.106.018501}.

\end{thebibliography}

%% ------------------------------------------------------------------------ %%
%
%  END ARTICLE
%
%% ------------------------------------------------------------------------ %%

\end{article}

\end{document}